# New symmetric families of silicon quantum dots and their conglomerates as a tunable source of photoluminescence in nanodevices

Pavel V. Avramov[1, 2*†], Dmitri G. Fedorov[3*], Pavel B. Sorokin[2, 4, 5], Leonid A. Chernozatonskii[4], and Mark S. Gordon[6]

[1] Takasaki-branch, Advanced Science Research Center, Japan Atomic Energy Agency, Takasaki, 370-1292, Japan

[2] L.V. Kirensky Institute of Physics SB RAS, 660036 Krasnoyarsk, Russian Federation

[3] RICS, National Institute of Advanced Industrial Science and Technology (AIST), 1-1-1 Umezono, Tsukuba, Ibaraki, 305-8568, Japan

[4] N.M. Emanuel Institute of Biochemical Physics of RAS, 119334 Moscow, Russian Federation

[5] Siberian Federal University, 79 Svobodniy av., Krasnoyarsk, 660041 Russia

[6] Ames National Laboratory/Department of Chemistry, Iowa State University, Ames, Iowa 50011, USA

* These authors contributed equally to this work.

† e-mail: avramov.pavel@jaea.go.jp

We propose a new variety of silicon quantum dots containing fullerene-derived hollows of nearly arbitrary symmetry. Conglomerate structures are designed by connecting the quantum dots through two kinds of junctions. The quantum confinement effect is investigated using semiempirical quantum-mechanical method. It is shown that within each family of quantum dots, the band gap and the stability are inversely proportional to the particle effective size. Quantum dots inherit a wide variety of structural and symmetry properties from their parent fullerenes. The conglomerates confine electrons like quasi-molecules with a peculiar electronic structure related to the junctions. Quantum dots and their conglomerates can host guest atoms in their hollows and therefore present a new promising type of tunable photoluminescent nanomaterials.

Silicon nanowires and quantum dots have recently attracted much experimental and theoretical interest.[1,2,3,4,5,6,7,8] Porous silicon and silicon nanocrystals (*nc*-Si) precipitated into a $SiO_2$ matrix are two examples of silicon-based nanoscale systems. Structural investigations show that *nc*-Si particles, having a wide variety of shapes (including quasispherical) and sizes, retain the diamond-like atomic structure of bulk silicon. Although the general crystal lattice type of these species is clear, their exact atomic structure remains unknown.

The *nc*-Si experimental photoluminescence (PL) spectra, obtained for samples synthesized under different conditions (see, for example Refs. 9,10), differ from each other very significantly. The PL excitation energies are closely related to the band gap, and the quantum confinement effect (QCE) appears as a band gap dependence on the maximum linear size $d$ of the *nc*-Si particles with $A+Cd^{-k}$ form, where $A$, $C$ and $k$ represent the sample-dependent parameters. Probably the pronounced distinctions in QCEs are caused by the unresolved differences in the atomic structure of the existing *nc*-Si types.

Several kinds of silicon nanowires and quantum dots have been proposed and studied theoretically. Most species have square or rectangular cross-sections, but some nanowires[11] of tetra-, penta- and hexagonal symmetry, as well as quantum dots of tetrahedral[12], icosahedral or truncated cubic symmetry[13] have been reported. The icosahedral moiety, having 12 pentagonal vertices and retaining the diamond-like atomic structure, is predicted to have the lowest energy per atom among all small size nanoparticles ($d \leq 5$ nm). The DFT and semi-empirical electronic structure calculations have been applied to study the QCE for structures with different symmetry[8,14,15,16] and no deviation from the typical inverse $d$ dependence has been reported.

The previous theoretical calculations have so far addressed only a limited number of perfect symmetrical types of silicon clusters and have not described all possible shapes of experimental structures. The basic structural tetrahedral units of the silicon lattice have the perfect structure with four <111> facets. The rich diversity of the shapes and sizes of the nanocrystalline silicon can be explained by the chemical binding between *nc*-Si cores at the tetrahedral facets, edges and vertices. In the simplest case, the combination of 20 tetrahedra results in the formation of the perfect icosahedral quantum dot structure.[13] However, there is no reason why other symmetric families of quantum dots (QDs) with different numbers of tetrahedra cannot be formed, and we speculate that such nanoclusters can partially explain the rich diversity of the *nc*-Si structures.

We now formulate a mathematical definition and a general recipe to generate new quantum dot structures following the Goldberg polyhedron[17] pattern: quantum dots with a desired number of pentagonal and hexagonal vertices can be designed by bringing together 20+*n* silicon tetrahedra (Fig. 1a, b) ($n \geq 4$ is some integer) of the same sizes through three equivalent <111> facets, and the remaining facets (one in each tetrahedron) form the surface of the resultant quantum dot. The inward vertices of these silicon tetrahedra form low fullerene-like $Si_{20+n}$ regions in the center (Fig. 1c), composed of 12 pentagons and *n*/2 hexagons.

The symmetry of a quantum dot (Fig. 1d) is a subgroup of the point group of the parent fullerene, whose pentagons and hexagons are attached by pentagonal/hexagonal channels formed by five/six tetrahedral edges, respectively. The external channel ends are in fact the pentagonal/hexagonal vertices of the quantum dot. The addition of silicon tetrahedra results in some structural tension due to deviations of chemical bonds involving Si from the perfect tetrahedral arrangement (see below). Because of this, only small fullerene-based structures with a high symmetry $Si_{24}$ ($D_{6d}$), $Si_{26}$ ($D_{3h}$), and $Si_{28}$ ($T_d$) (Fig 2) were used in this work as basic units to produce stable QDs; one can, however, use any other symmetry (subgroups of $I_h$ or $D_{6d}$). The spacious hollow (roughly, 5.1x7.7x7.7 Å$^3$) for one quantum dot with the basic $Si_{24}$ fullerene-like core ($D_{6d}$) is presented in Fig. 1c. The hollow is large enough to hold one or several guest ions atoms or molecules forming an endohedral complex.

A practical way to design a Goldberg type quantum dot is the following: on the top of the fullerene core (Fig. 1c), a second layer of 20+*n* atoms is added. The atoms of the third layer are connected with each other by the atoms forming the surface of the resulting QD with 12 pentagonal and *n/2* hexagonal vertices. Using the same procedure one can add several silicon layers forming a QD of the desired size and symmetry. All quantum dots with *L* silicon layers built upon the fullerene-like core with *m* atoms having the point group *G,* can be compactly classified under the notation of $L_m^G$. For example, the $2_{20}^{\mathrm{I}_h}$ and $3_{24}^{\mathrm{D}_{6d}}$ symbols denote the two- and three-layered icosahedral ($Si_{100}H_{60}$) and hexagonal ($Si_{336}H_{144}$) structures, respectively.

It is possible to introduce a stoicheometric formula for the species with *L* silicon layers. The number of silicon atoms in each individual silicon layer *l* is equal to $m*l^2$, where *m* is the number of silicon atoms of the core. The number of silicon atoms $N_{\mathrm{Si}}$ for a given number of layers *L* is equal to $N_{\mathrm{Si}} = m\sum_{l=1}^{L} l^2 = mL(L+1)(2L+1)/6$ . The number of hydrogen atoms saturating the dangling bonds for the silicon layers is given by $N_{\mathrm{H}} = m\sum_{l=1}^{L} l = mL(L+1)/2$. E.g., the $3_{24}^{\mathrm{D}_{6d}}$ dot has *m*=24, *L*=3, thus $N_{\mathrm{Si}} = 336$ and $N_{\mathrm{H}} = 144$.

To perform a systematic comparative study of the electronic properties of silicon QDs, we also considered a number of previously reported structures, closely related to bulk silicon; cubic (denoted by $\mathrm{C}_N^{\mathrm{O}_h}$, where the central symbol C is for cubic, the superscript denotes the point group of the basic unit and the subscript shows the number of silicon atoms *N* in the whole structure), 14-facet truncated cubic ($\dot{\mathrm{C}}_N^{\mathrm{O}_h}$, where the single dot denotes the truncation of vertices), 26-facet truncated cubic ($\ddot{\mathrm{C}}_N^{\mathrm{O}_h}$, two dots denote the truncation of the vertices and the edges),[13] octahedral ($\mathrm{O}_N^{\mathrm{O}_h}$), truncated octahedral ($\dot{\mathrm{O}}_N^{\mathrm{O}_h}$) and tetrahedral ($\mathrm{T}_N^{\mathrm{T}_d}$) quantum dots. The tetrahedral and octahedral QDs have 4 and 8 <111> facets, respectively. All optimized geometries are provided in Supplementary Materials.

Junctions of pentagonal or hexagonal symmetry can be made by cutting off two vertices of a pair of proposed quantum dots and connecting the truncated structures through the resulting cross-sections into one conglomerate structure. Let us denote the interfaces through hexagonal and pentagonal vertices by empty ($\circ$) and solid ($\bullet$) circles, respectively. Such junctions keep the tetrahedral nature of all silicon atoms constituting the structures. We have designed several conglomerates composed of two, three, four or six QDs of different symmetry, size and shape (linear, bent and arrange in a circle $c-\left(2_{26}^{D_{3h}} \circ 2_{26}^{D_{3h}}\right)_3$ structures, $c$ stands for cyclic, Fig. 2). Similar linear structures composed of several icosahedral quantum dots, have been obtained in classical MD simulations by freezing the silicon melt in a thin (1.36 nm) nanopore.[18] In multilayered quantum dots ($L$≥3), the junctions can be modified by the way the vertexes are cut and connected with each other (for details, see α, β and γ forms of $3_{20}^{I_h} \bullet 3_{20}^{I_h}$ structure in Supplementary materials).

The formation of the Goldberg QDs affects the perfect tetrahedral structure of silicon. Using the semi-empirical AM1 molecular orbital method (see the last section of the paper) the Si-Si distance in the center of the largest silicon tetrahedron $Si_{281}H_{172}$ cluster ($T_{281}^{T_d}$, a good approximation to the bulk silicon) is predicted to be 2.340 Å. The smallest $2_{20}^{I_h}$ structure has three nonequivalent Si-Si bonds with predicted distances of 2.332, 2.321 and 2.378 Å; the diameter of the central hollow is 6.536 Å. The deviation from the perfect terahedral values of the bond angles reaches ±2.40˚. The central hollow region in $3_{20}^{I_h}$ has a diameter of 6.482 Å and larger deviations of the bond lengths (2.304-2.366 Å) and angles (106.56-111.9˚).

The $2_{24}^{D_{6d}}$ structure has larger structural distortions that lead to strain, because of the $D_{6d}$ hollow with the smallest and largest dimensions of 5.093 and 7.679 Å, respectively (Fig. 1). Both hexagons of the $Si_{24}$ core keep their perfect structure with 120˚ bond angles and 2.375 Å Si-Si bond lengths. Because of the structural deformation, other angles around the core are smaller (106.22˚, 108.08˚ and 108.48˚) than the perfect tetrahedral value (109.5˚). The bong lengths and angles of other silicon layers vary from 2.342 Å to 2.375 Å and from 97.58˚ to 114.52˚, respectively.

The conglomerate structures like $L_{20}^{I_h} \bullet L_{20}^{I_h}$ have a slight (~1%) deviation of the Si-Si bond lengths in the interface region with respect to the parent ( $L_{20}^{I_h}$ ) structures. The symmetric interfaces ( $2_{24}^{D_{6d}} \circ 2_{24}^{D_{6d}}$ , $2_{26}^{D_{3h}} \circ 2_{26}^{D_{3h}}$ and $2_{28}^{T_d} \circ 2_{28}^{T_d}$ ) have mirror symmetry interface regions and a similar slight deviation of bond lengths and angles. The asymmetric interfaces, like $2_{24}^{D_{6d}} \circ 2_{26}^{D_{3h}}$ , $2_{28}^{T_d} \bullet 2_{26}^{D_{3h}}$ etc., have irregular complex atomic structures with unexpected distortions of the interface regions.

The AM1 energetic stability of all systems (see Tables 1 and 2 of Supplementary Materials) is plotted in Fig. 3a. For comparison, the Si atom energy ($^3$P term) at the same level of theory is -1821.81 kcal/mol. Thus, within our definition of the averaged atomic energy in clusters, Si atoms in quantum dots and their conglomerates are predicted to be stabilized by about 80-100 kcal/mol per atom. In the region of 1.2–3.5 nm, the icosahedral quantum dots are the most stable structures, which is in agreement with the earlier density functional theory (DFT) results.[13] The $\ddot{C}_N^{O_h}$ type of structures is second in energy, which was also reported in the DFT study.[15] Some non-monotonic behavior of the $\ddot{C}_N^{O_h}$ and $\dot{C}_N^{O_h}$ curves can be explained by the formation of silicon dimers at the <100> surfaces. [15,15,15] In the 1.5-2.24 nm region, the proposed structures of the $D_{6d}$, $D_{3h}$ and $T_d$ cores have a higher stability than the truncated octahedral structures.

The conglomerate stabilities are higher than those of their single parent quantum dots. Excluding the mixed kinds, the stability increases in the inverse proportion to the linear size of the particles. The asymmetrical types of conglomerates (e.g., $2_{20}^{I_h} \bullet 2_{24}^{D_{6d}}$ or $2_{26}^{D_{3h}} \circ 2_{28}^{T_d}$ ) composed of several quantum dots of different symmetry have a higher stability than their building blocks ( $2_{20}^{I_h}$ etc), and they, if made from $2_{20}^{I_h}$, have a lower stability than the corresponding symmetric conglomerates. The observed increase in the energetic stability with the size is an important point practically in designing and producing nanodevices.

Regardless of the symmetry, the QCE dependence is observed within each family (Fig. 3b). The QCE curves have different inclines and limiting values and sometimes cross each other, as can be seen for the $2_{24}^{D_{6d}}$ and $2_{26}^{D_{3h}}$ -based objects. The presence of junctions in conglomerates changes the QCE, e.g., the linear $2_{20}^{I_h} \bullet 2_{20}^{I_h}$ structures reveal a noticeable decrease of the QCE slope in comparison with the corresponding single quantum dots. In general, the conglomerate structures have a somewhat larger band gap than the corresponding single QDs of the same size, caused by the effect of the junction upon the quantum confinement, which can be understood in terms of the quasi-molecular representation of conglomerates, see below. The QCE dependence of the symmetric conglomerates ( $2_{24}^{D_{6d}}$ , $2_{26}^{D_{3h}}$ and $2_{28}^{T_d}$ ) is similar to that of the $2_{20}^{I_h}$ -based structures, but with smaller values of the band gap energies and curve inclines.

The family of asymmetrical conglomerates ( $2_{20}^{I_h} \bullet 2_{24}^{D_{6d}}$ , $2_{20}^{I_h} \bullet 2_{28}^{T_d}$ , $2_{20}^{I_h} \bullet 2_{26}^{D_{3h}}$ , $2_{24}^{D_{6d}} \circ 2_{26}^{D_{3h}}$ , $2_{24}^{D_{6d}} \circ 2_{28}^{T_d}$ and $2_{26}^{D_{3h}} \circ 2_{28}^{T_d}$ ) shows a formal destruction of the typical QCE and a maximum at 2.5 nm. However, this departure from the typical dependence occurs because the set of data corresponds to structures of different types. The same apparent nonsystematic behavior would be seen if one plotted a mixture of simple QDs of different types on a single curve.

Due to the nearly spherical shapes, the excitons in quantum dots can be described as quasi-atomic states with the symmetry labeling in the atomic group $K_h$.[1] The multiglobular objects introduced in this work can be thought of as quasi-molecular objects (Fig. 2). The dots consisting of two globules (e.g., $2_{20}^{I_h} \bullet 2_{20}^{I_h}$ ) are quasi-diatomics, for which excitons can be labeled in $D_{\infty h}$. Some quasi-molecular conglomerates retain the full symmetry, even with the detailed atomic structure taken into consideration (e.g., the bent structure of $2_{28}^{T_d} \circ 2_{28}^{T_d} \circ 2_{28}^{T_d} \circ 2_{28}^{T_d}$ in $C_{2h}$ symmetry), whereas linear structures like $2_{20}^{I_h} \bullet 2_{20}^{I_h} \bullet 2_{20}^{I_h} \bullet 2_{20}^{I_h}$ and $2_{24}^{D_{6d}} \circ 2_{24}^{D_{6d}} \circ 2_{24}^{D_{6d}} \circ 2_{24}^{D_{6d}}$ with the $D_{5h}$ and $D_{6h}$ symmetry, respectively, are not much below the full group of $D_{\infty h}$.

The highest occupied molecular orbitals (HOMO), Fig. 2, describe the Si-Si $\sigma$-bonds delocalized over many pairs of atoms. The sign alternation between neighbors is commonly seen in all structures (see the intermingled blue and red orbital droplets for $3_{24}^{D_{6d}}$), with some orbital clouds assembled in larger nebular aggregates between the silicon layers, as can be seen for $2_{20}^{I_h}$. The HOMO localization in conglomerates is quite interesting: whereas for the symmetric ("$D_{\infty h}$") type much density remains inside the globules ($2_{20}^{I_h} \bullet 2_{20}^{I_h} \bullet 2_{20}^{I_h}$ and $\alpha\text{-}c-\left(2_{26}^{D_{3h}} \circ 2_{26}^{D_{3h}}\right)_3$) and pronounced nodes between globules are often found, for the asymmetric kind ("$C_{\infty v}$"), HOMOs are strongly localized in the interglobule junction area (e.g., $2_{24}^{D_{6d}} \circ 2_{28}^{T_d}$). The reason for this behavior is the increased tension between globules, which is expressed in the higher orbital energies, pushing one junction MO to become the HOMO localized in the interglobule junction area ($2_{28}^{T_d} \circ 2_{28}^{T_d} \circ 2_{28}^{T_d} \circ 2_{28}^{T_d}$) or on the inner ring for the torus $c-\left(2_{26}^{D_{3h}} \circ 2_{26}^{D_{3h}}\right)_3$.

The view of the multiglobular structures as quasi-polyatomics can be further developed by considering the orbitals in the junction area to be orbitals describing the chemical bonding between quasi-atoms (globules). The HOMO for $2_{24}^{D_{6d}} \circ 2_{28}^{T_d}$ shows the opposite phases between the globules, and thus one can speculate that this HOMO actually looks like the occupied antibonding orbital between quasi-atoms, whereas for $2_{28}^{T_d} \circ 2_{28}^{T_d} \circ 2_{28}^{T_d} \circ 2_{28}^{T_d}$ some shared electron density between globules can be seen, purporting a certain bonding character.

In this work, based on the fullerene-like central silicon cores, we proposed and systematically classified new families of silicon quantum dots of adjustable symmetry. The central hollows can accommodate one or more guest atoms or molecules, and the interplay of several hosts in conglomerates can lead to promising novel types of nanodevices. All structures are stable minima on the energy surface, display an increased stability with the linear size and obey the same relation for the quantum confinement effect. The detailed structural information can be used to aid the analysis of the experimental data, e.g., if mass or Raman spectra are measured, the structures can be matched with the symmetries and chemical formulas we have provided here.

Each family of nanoparticles features a unique band gap dependence upon the linear size. By varying the symmetry and the size one can make light emitters of a desired wavelength or prepare a mixture of particles producing composite colors, such as cyan or white. The low fullerenes determine the symmetric properties of the hollow in the nanodots and their conglomerates; this is especially important for the endohedral complexes when the symmetry of the hollow directly controls the electronic structure splitting of the guest. The combination of all predicted structural, symmetric and electronic properties of the new silicon quantum dots and their conglomerates has the potential to open a new field of applications in nanoscience and nanotechnology.

**Methods**

To study the structure and electronic properties we used the semi-empirical Austin Model 1 (AM1)[19] based on the modified neglect of diatomic overlap (MNDO)[20] approximation, which has been successfully employed previously to study the atomic and electronic structures of Si nanoclusters saturated by hydrogen atoms.[8,12,21] AM1 is generally thought to produce very good structures and to systematically overestimate the band gap in silicon clusters. For single quantum dots cut out from bulk silicon, we fitted the band gap to the form of $A + Cd^{-1}$ and obtained its value at the infinite size *d,* equal to *A*=6.070 eV. The experimental value in solid crystal silicon is 1.16 eV,[23] thus we assumed that AM1 band gap overestimates experiment by 4.91 eV and this value was subtracted from all AM1 band gap values in Fig. 3b and Tables 1, 2 of Supplementary Materials. The atomic structure optimization was carried out until the RMS gradient became smaller than 0.05 kcal/mol/Å. The point groups were determined based on strict criteria suitable for quantum-mechanical orbital labeling; however, most structures possess a much higher symmetry (also listed in Supplemental Materials) within a minor distortion allowance (in the RMS terms, typically about 0.01 Å).

To study the stability, the energy per silicon atom for all systems was calculated. Due to the nearly perfect tetrahedral character of these Si clusters, the energies of hydrogen atoms ($E_H$) can be assumed to be close to a constant, which we calculated from the difference of the energies of several $Si_mH_n$ and $Si_mH_{n-2}$ tetrahedral clusters to be equal to -356.86 kcal/mol. Consequently, the averaged energy per Si atom in a $Si_mH_n$ structure was defined as $E_{Si} = \frac{E - nE_H}{m}$, where $E$ is the semiempirical energy of $Si_mH_n$.

## Acknowledgments

This work was partially supported by the project “Material design with New Functions Employing Energetic Beams” and JAEA Research fellowship (PVA), by Russian Fund of Basic Researches (grant n.05-02-17443), grants from the US Department of Energy via the Ames Laboratory and the Air Force Office of Scientific Research, grant of Deutsche Forschungsgemeinschaft and Russian Academy of Sciences, No. 436 RUS 113/785 (LAC). PVA also acknowledges the members of “Research Group for Atomic-scale Control for Novel Materials under Extreme Conditions” Prof. H. Naramoto, Prof. Y. Maeda, Dr. K. Narumi and Dr. S. Sakai for fruitful discussions and hospitality.

**FIGURE LEGENDS**

Fig. 1. Construction of a typical representative ($2_{24}^{D_{6d}}$) of the new variety of quantum dots with the inner $Si_{24}$ core of $D_{6d}$ symmetry. (a) 24 silicon tetrahedra $Si_5$ collapsed into a Goldberg polyhedron. (b) Silicon tetrahedra brought close with a gap between them, forming the $2_{24}^{D_{6d}}$ quantum dot with 24*5=120 atoms. The triangular facets of the tetrahedra facing the surface are shown in green. Note that there are no chemical bonds between the vertices of the tetradehra, and their edges are shown for geometrical reasons. Two typical vertexes with hexagonal and pentagonal facets involving actual Si-Si bonds are shown in magenta. (c) The $Si_{24}$ hollow inside the quantum dot $2_{24}^{D_{6d}}$ has a low fullerene structure, with the smallest and largest linear dimensions shown. (d) The fully optimized structure of the quantum dot with a hollow (shown in red). The silicon atoms are depicted in layers 1, 2 and 3 in red, green and gray colors, respectively. Hydrogen atoms terminating unsaturated surface bonds are not shown.

Fig. 2. The proposed silicon structure highest occupied molecular orbitals are shown immediately below each structure, for a selected set of silicon quantum dots (upper part) and their conglomerates (lower part). The two phases of the orbitals are shown in blue and red. The hollows inside the structures are shown in red. The symbols of each structure describe the number of silicon layers, the size and symmetry of the central hollow (see main text). Single globe quantum dots (top 4 objects) are nearly spherical structures and thus atom-like in terms of the electron confinement; the lower 4 conglomerate quantum dots are multiglobular molecule-like structures.

Fig. 3. a) Energetic stability (per Si atom) of the complex silicon nanoclusters vs the effective size $d$ (Å) shown as: black line with empty diamonds ($L_{26}^{D_{3h}}$), gray line with filled squares ($L_{28}^{T_d}$), dark cyan line with empty triangles ($L_{24}^{D_{6d}}$), cyan line with filled circles ($L_{20}^{I_h}$), black line with filled circles ($T_N^{T_d}$), red line with empty triangles (octahedral, $O_N^{O_h}$), filled squares (truncated octahedral, $\dot{O}_N^{O_h}$), empty diamonds (cubic, $C_N^{O_h}$), empty hexagons

(truncated cubic, $\dot{C}_N^{O_h}$), filled triangles (doubly truncated cubic, $\ddot{C}_N^{O_h}$), green line with empty triangles ($\left(2_{20}^{I_h}\right)_n$), brown line with filled squares ($\left(2_{24}^{D_{6d}}\right)_n$), green line with empty diamonds ($\left(2_{26}^{D_{3h}}\right)_n$ $n$ = 2, 3 and 4), dark yellow line with filled triangles (torus, $\left(2_{26}^{D_{3h}}\right)_6$), blue line with empty hexagons ($\left(2_{28}^{T_d}\right)_n$) and violet line with points (asymmetric conglomerates).

b) Band gap (quantum confinement) vs the effective size $d$ (Å), shown as black line (typical experimental[7] QCE of porous silicon), black line with empty diamonds ($L_{26}^{D_{3h}}$), dark cyan line with empty triangles ($L_{24}^{D_{6d}}$), cyan line with filled circles ($L_{20}^{I_h}$), black line with filled circles ($T_N^{T_d}$), green line with empty triangles ($\left(2_{20}^{I_h}\right)_n$), brown line with filled squares ($\left(2_{24}^{D_{6d}}\right)_n$) and green line with empty diamonds ($\left(2_{26}^{D_{3h}}\right)_n$).

Fig. 1

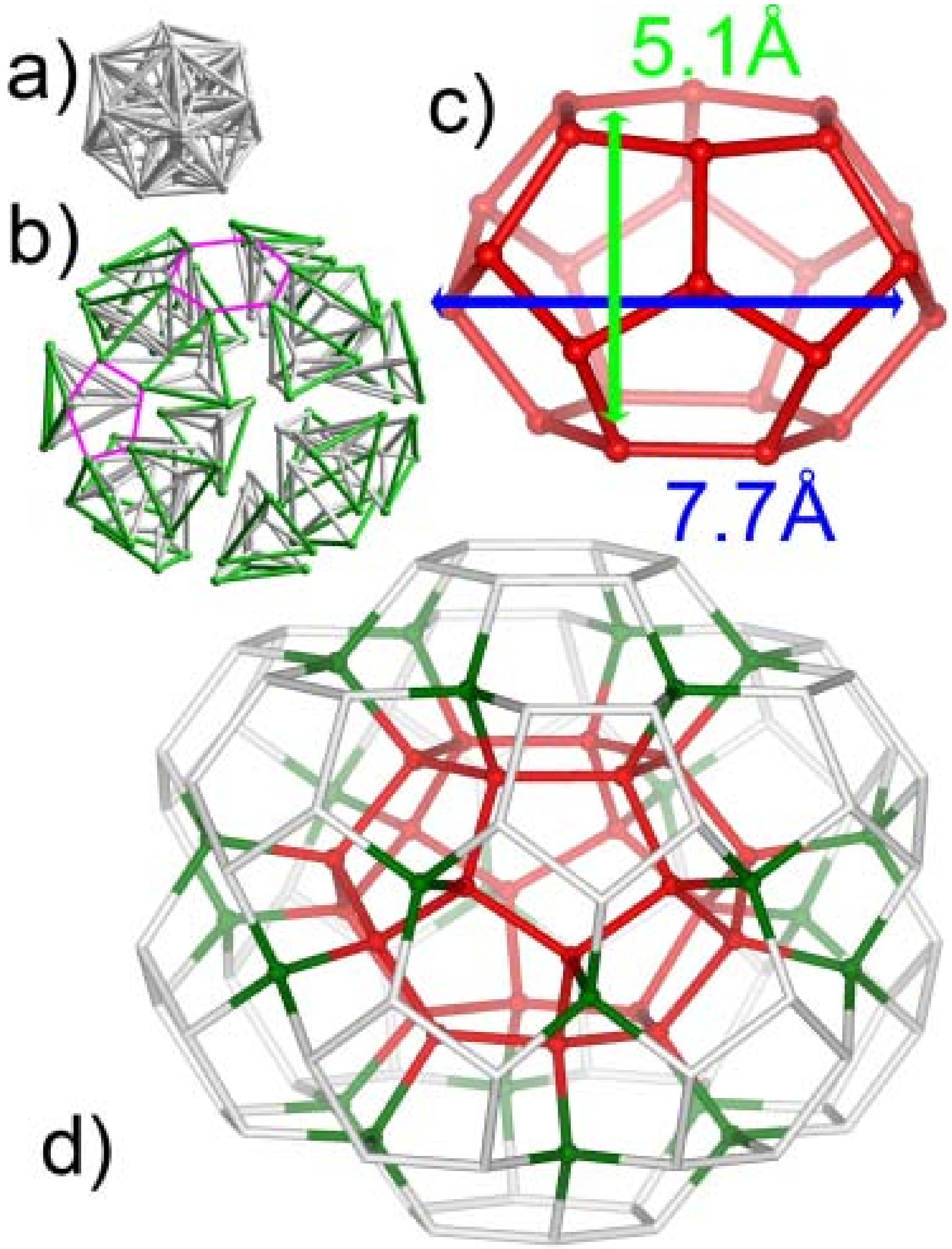

Fig 2.

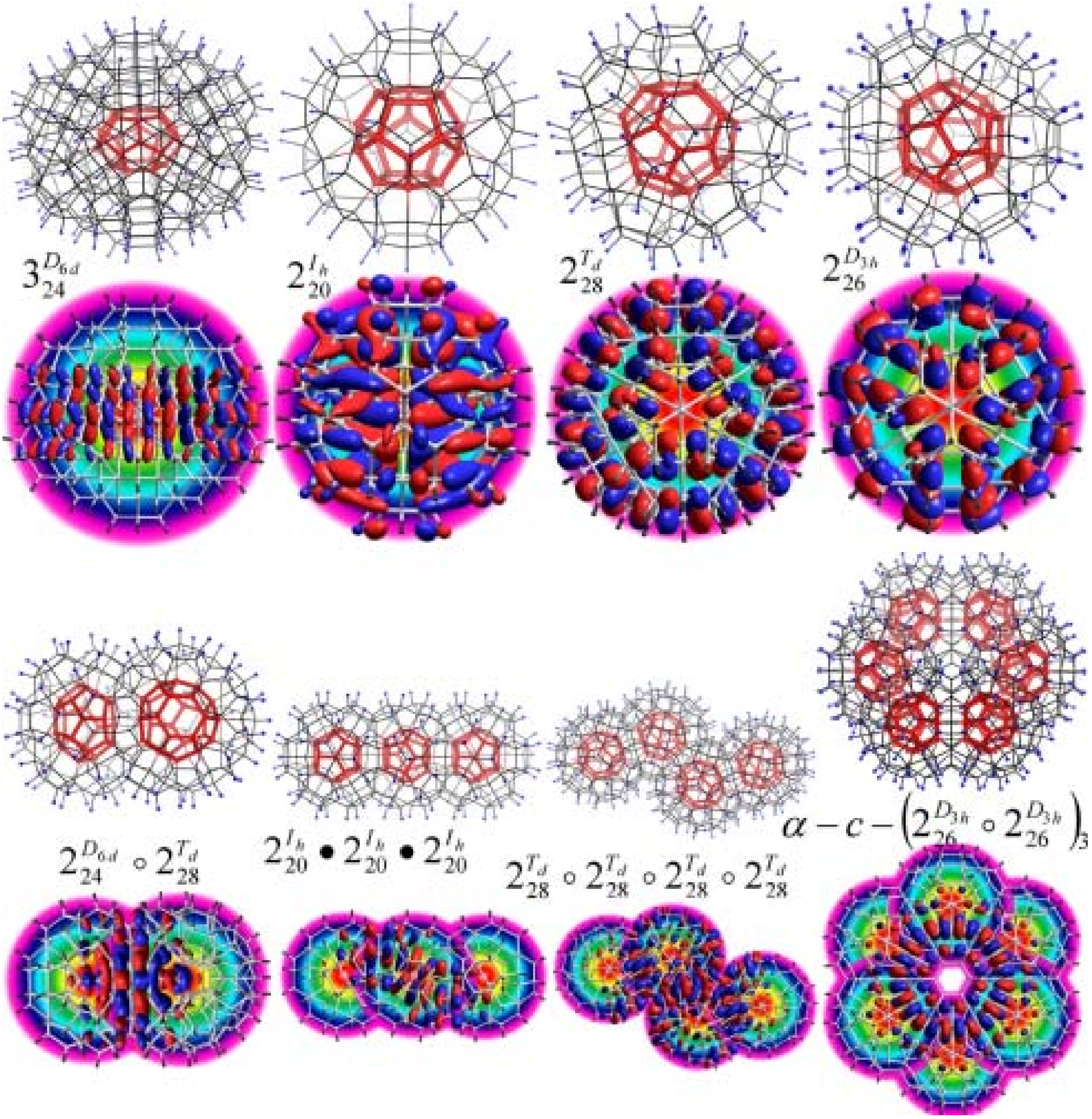
$3_{24}^{D_{6d}}$
$2_{20}^{I_h}$
$2_{28}^{T_d}$
$2_{26}^{D_{3h}}$
$\alpha - c - \left(2_{26}^{D_{3h}} \circ 2_{26}^{D_{3h}}\right)_3$
$2_{24}^{D_{6d}} \circ 2_{28}^{T_d}$
$2_{20}^{I_h} \bullet 2_{20}^{I_h} \bullet 2_{20}^{I_h}$
$2_{28}^{T_d} \circ 2_{28}^{T_d} \circ 2_{28}^{T_d} \circ 2_{28}^{T_d}$

Fig. 3

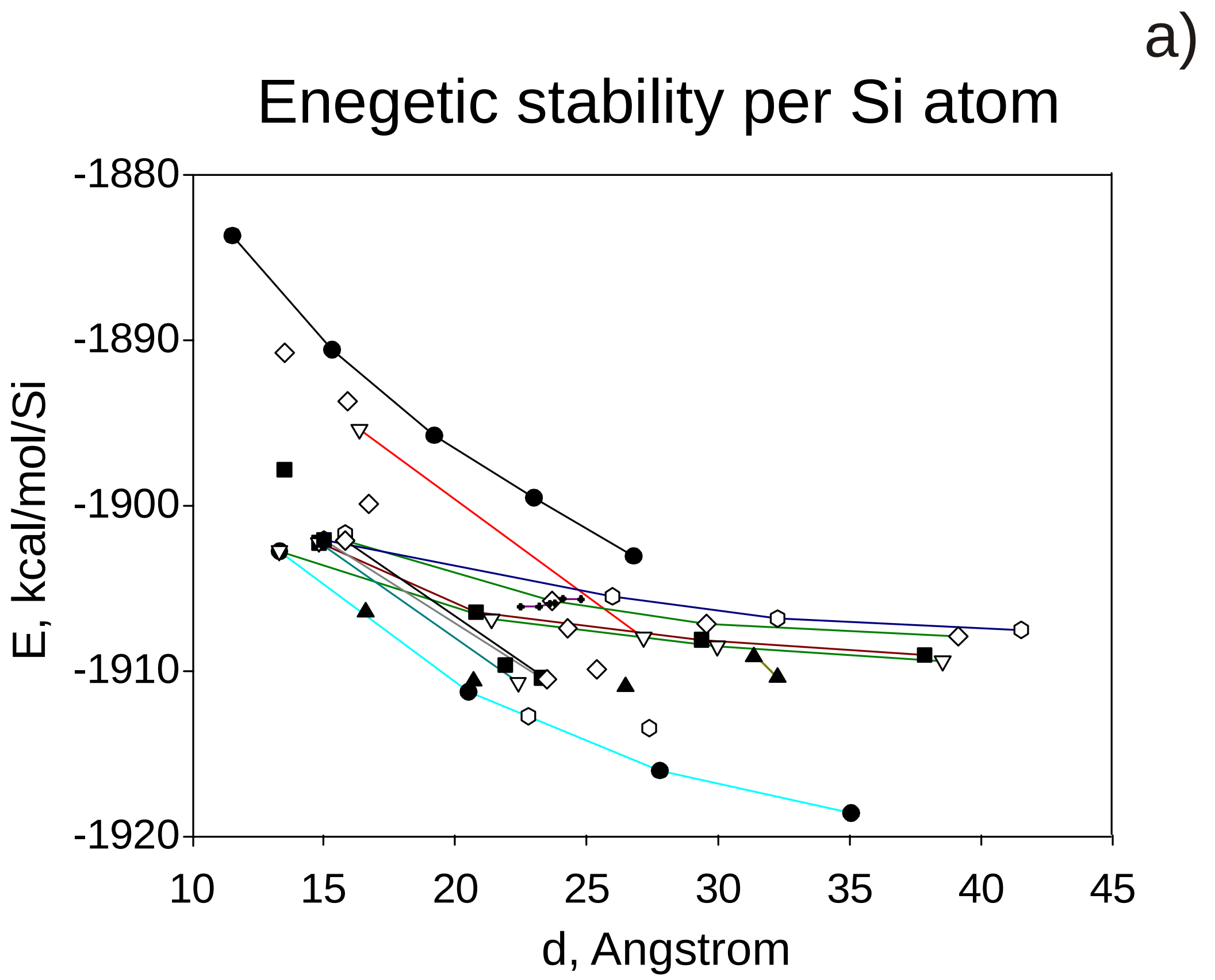

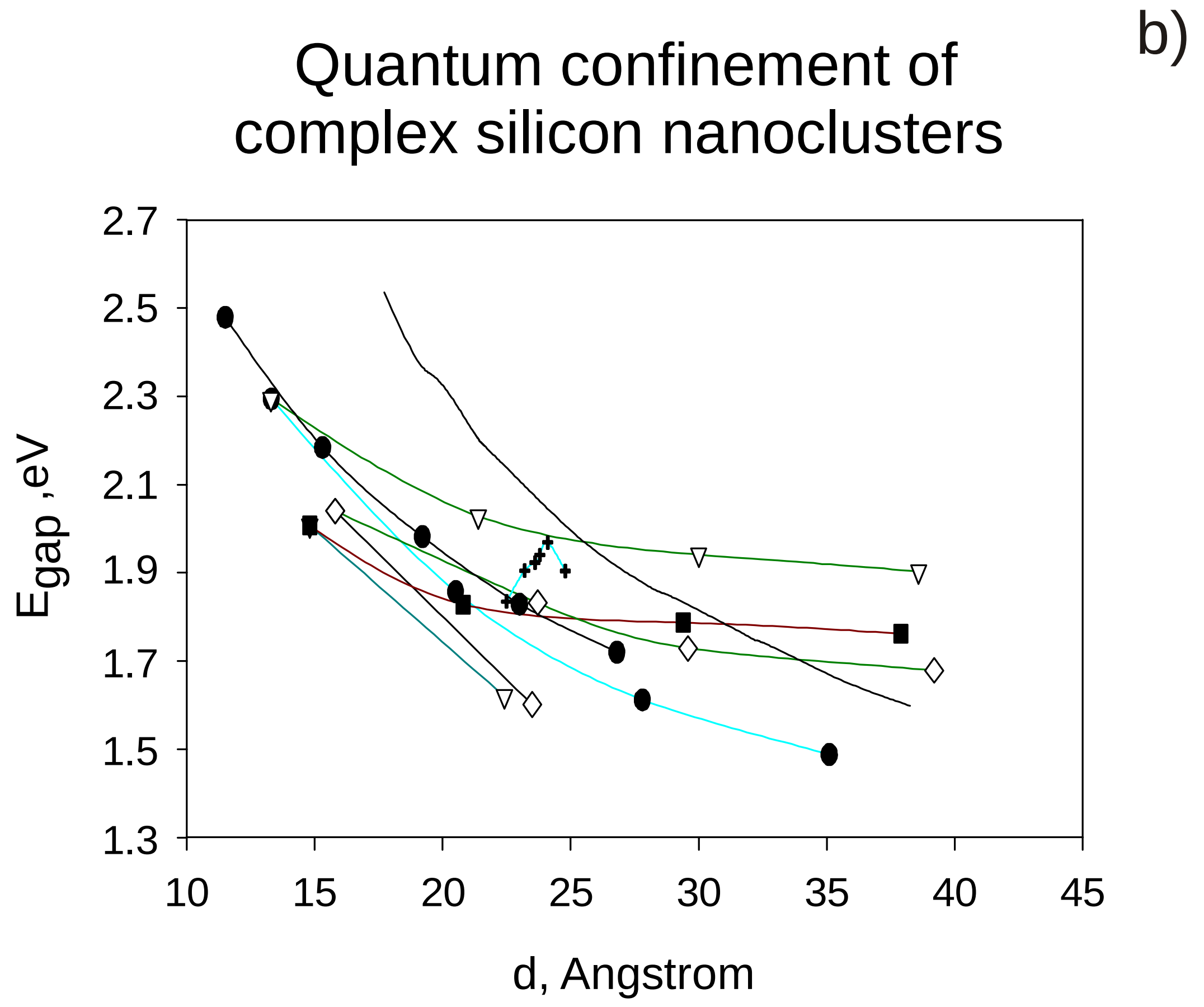

Table 1. Properties of single silicon quantum dots (the energy $E_{Si}$ per Si atom shows stability).

| Name | Formula | Symmetry[a] | min size, nm | max size, nm | HOMO, LUMO | Band gap, eV[b] | $E_{Si}$, kcal/mol |
|---|---|---|---|---|---|---|---|
| Trigonal, hexagonal and tetrahedral (fullerene-derived) | | | | | | | |
| $2_{26}^{D_{3h}}$ | $Si_{130}H_{78}$ | $C_{3h}/D_{3h}$ | 1.58 | 1.58 | A”,E’ | 2.028 | -1902.11 |
| $2_{24}^{D_{6d}}$ | $Si_{120}H_{72}$ | $D_{6d}$ | 1.48 | 1.48 | $B_1$,$B_2$ | 1.997 | -1902.23 |
| $2_{28}^{T_d}$ | $Si_{140}H_{84}$ | $C_1/C_{3v}$ | 1.50 | 1.50 | A,A | 2.136 | -1902.05 |
| $3_{26}^{D_{3h}}$ | $Si_{364}H_{156}$ | $C_{3h}/D_{3h}$ | 2.35 | 2.35 | A",E' | 1.159 | -1910.50 |
| $3_{24}^{D_{6d}}$ | $Si_{336}H_{144}$ | $D_{6d}$ | 2.24 | 2.24 | $B_1$,$B_2$ | 1.611 | -1910.67 |
| $3_{28}^{T_d}$ | $Si_{392}H_{168}$ | $C_1/S_4$ | 2.33 | 2.33 | A,A | 1.686 | -1910.40 |
| Icosahedral | | | | | | | |
| $2_{20}^{I_h}$ | $Si_{100}H_{60}$ | $C_i/T_h$ | 1.33 | 1.33 | $A_g$,$A_u$ | 2.284 | -1902.74 |
| $3_{20}^{I_h}$ | $Si_{280}H_{120}$ | $C_i/T_h$ | 2.05 | 2.05 | $A_g$,$A_u$ | 1.847 | -1911.26 |
| $4_{20}^{I_h}$ | $Si_{600}H_{200}$ | $C_1/T_h$ | 2.78 | 2.78 | A,A | 1.602 | -1916.01 |
| $5_{20}^{I_h}$ | $Si_{1100}H_{300}$ | $C_1/C_5$ | 3.51 | 3.51 | A,A | 1.457 | -1918.58 |
| Tetrahedral | | | | | | | |
| $T_{51}^{T_d}$ | $Si_{51}H_{52}$ | $C_1/D_{2d}$ | 0.94 | 1.15 | A,A | 2.469 | -1883.66 |
| $T_{87}^{T_d}$ | $Si_{87}H_{76}$ | $C_1/C_{3v}$ | 1.35 | 1.53 | A,A | 2.174 | -1890.57 |
| $T_{136}^{T_d}$ | $Si_{136}H_{104}$ | $C_1/S_4$ | 1.64 | 1.92 | A,A | 1.972 | -1895.75 |
| $T_{201}^{T_d}$ | $Si_{201}H_{138}$ | $C_1/C_{3v}$ | 1.97 | 2.30 | A,A | 1.819 | -1899.51 |
| $T_{281}^{T_d}$ | $Si_{281}H_{172}$ | $C_1/T_d$ | 2.28 | 2.68 | A,A | 1.710 | -1903.04 |
| Octahedral | | | | | | | |
| $O_{84}^{O_h}$ | $Si_{84}H_{64}$ | $C_1/D_{2d}$ | 1.64 | 1.64 | A,A | 2.239 | -1895.39 |
| $O_{286}^{O_h}$ | $Si_{286}H_{144}$ | $C_1/C_3$ | 2.72 | 2.72 | A,A | 1.709 | -1907.97 |
| Truncated Octahedral | | | | | | | |
| $\dot{O}_{80}^{O_h}$ | $Si_{80}H_{56}$ | $D_{2d}$ | 1.35 | 1.35 | E,$B_2$ | 2.190 | -1897.81 |
| $\dot{O}_{264}^{O_h}$ | $Si_{268}H_{124}$ | $C_{2v}$ | 2.19 | 2.19 | $B_1$,$B_1$ | 1.697 | -1909.63 |
| Cubic | | | | | | | |
| $C_{75}^{O_h}$ | $Si_{75}H_{64}$ | $C_1$ | 1.13 | 1.35 | A,A | 2.148 | -1890.74 |
| $C_{88}^{O_h}$ | $Si_{88}H_{70}$ | $C_1$ | 0.96 | 1.59 | A,A | 2.118 | -1893.68 |
| $C_{139}^{O_h}$ | $Si_{139}H_{92}$ | $C_1$ | 1.54 | 1.67 | A,A | 1.877 | -1899.90 |
| $C_{280}^{O_h}$ | $Si_{280}H_{140}$ | $C_1$ | 1.59 | 2.43 | A,A | 1.644 | -1907.41 |
| $C_{368}^{O_h}$ | $Si_{368}H_{164}$ | $C_s$ | 1.88 | 2.54 | A’,A’ | 1.555 | -1909.91 |
| Truncated cubic | | | | | | | |
| $\dot{C}_{123}^{O_h}$ | $Si_{123}H_{76}$ | $C_1$ | 0.96 | 1.58 | A,A | 1.890 | -1901.68 |
| $\dot{C}_{323}^{O_h}$ | $Si_{323}H_{124}$ | $C_1/T$ | 1.06 | 2.28 | A,A | 1.601 | -1912.71 |
| $\dot{C}_{537}^{O_h}$ | $Si_{537}H_{404}$ | $C_1$ | 2.01 | 2.74 | A,A | 1.482 | -1913.43 |
| Doubly truncated cubic | | | | | | | |
| $\ddot{C}_{147}^{O_h}$ | $Si_{147}H_{76}$ | $C_1/T_d$ | 1.37 | 1.66 | A,A | 1.904 | -1906.42 |
| $\ddot{C}^{O_h}_{287}$ | $Si_{287}H_{124}$ | $C_1/S_4$ | 1.92 | 2.07 | A,A | 1.641 | -1910.59 |
| $\ddot{C}^{O_h}_{465}$ | $Si_{465}H_{204}$ | $C_1$ | 2.17 | 2.65 | A,A | 1.514 | -1910.92 |

[a,b] footnotes: see below Table 2.

Table 2. Properties of conglomerate quantum dots (the energy $E_{Si}$ per Si atom shows stability).

| Name | Formula | Symmetry[a] | Min size, nm | Max size, nm | HOMO, LUMO | Band gap, eV[b] | $E_{Si}$, kcal/mol |
|---|---|---|---|---|---|---|---|
| $\left(2^{I_h}_{20}\right)_n$ family | | | | | | | |
| $2^{I_h}_{20} \bullet 2^{I_h}_{20}$ | $Si_{175}H_{90}$ | $D_5/D_{5h}$ | 1.33 | 2.14 | $E_2,A_2$ | 2.018 | -1906.85 |
| $2^{I_h}_{20} \bullet 2^{I_h}_{20} \bullet 2^{I_h}_{20}$ | $Si_{250}H_{120}$ | $C_1/D_{5d}$ | 1.33 | 3.00 | A,A | 1.931 | -1908.49 |
| $2^{I_h}_{20} \bullet 2^{I_h}_{20} \bullet 2^{I_h}_{20} \bullet 2^{I_h}_{20}$ | $Si_{325}H_{150}$ | $C_1/D_{5h}$ | 1.33 | 3.86 | A,A | 1.893 | -1909.38 |
| $2^{I_h}_{20} \bullet 3^{I_h}_{20}$ family | | | | | | | |
| $2^{I_h}_{20} \bullet 3^{I_h}_{20}$ | $Si_{355}H_{150}$ | $C_1/C_{5v}$ | 1.33 | 2.88 | A,A | 1.773 | -1911.52 |
| $2^{I_h}_{20} \bullet 3^{I_h}_{20} \bullet 2^{I_h}_{20}$ | $Si_{430}H_{180}$ | $C_1/D_5$ | 1.33 | 3.77 | A,A | 1.735 | -1911.68 |
| $\left(3^{I_h}_{20}\right)_n$ family | | | | | | | |
| $\alpha$-$3^{I_h}_{20} \bullet 3^{I_h}_{20}$ | $Si_{535}H_{210}$ | $C_1/D_{5h}$ | 2.05 | 3.62 | A,A | 1.652 | -1913.03 |
| $\beta$-$3^{I_h}_{20} \bullet 3^{I_h}_{20}$ | $Si_{550}H_{210}$ | $C_s/D_{5h}$ | 2.05 | 3.60 | A",A" | 1.543 | -1913.60 |
| $\gamma$-$3^{I_h}_{20} \bullet 3^{I_h}_{20}$ | $Si_{490}H_{180}$ | $C_s/D_{5h}$ | 2.05 | 3.22 | A",A" | 1.655 | -1914.18 |
| $3^{I_h}_{20} \bullet 3^{I_h}_{20} \bullet 3^{I_h}_{20}$ | $Si_{700}H_{240}$ | $C_1/D_{5d}$ | 2.05 | 4.42 | A,A | 1.613 | -1914.89 |
| $\left(2^{D_{6d}}_{24}\right)_n$ family | | | | | | | |
| $2^{D_{6d}}_{24} \circ 2^{D_{6d}}_{24}$ | $Si_{210}H_{108}$ | $D_{6h}$ | 1.48 | 2.08 | $B_{1g},A_{2u}$ | 1.818 | -1906.42 |
| $2^{D_{6d}}_{24} \circ 2^{D_{6d}}_{24} \circ 2^{D_{6d}}_{24}$ | $Si_{300}H_{144}$ | $C_1/D_{6d}$ | 1.48 | 2.94 | A,A | 1.777 | -1908.10 |
| $2^{D_{6d}}_{24} \circ 2^{D_{6d}}_{24} \circ 2^{D_{6d}}_{24} \circ 2^{D_{6d}}_{24}$ | $Si_{390}H_{180}$ | $C_1/D_{6h}$ | 1.48 | 3.79 | A,A | 1.752 | -1909.00 |
| $\left(2^{D_{3h}}_{26}\right)_n$ family | | | | | | | |
| $2^{D_{3h}}_{26} \circ 2^{D_{3h}}_{26}$ | $Si_{230}H_{120}$ | $C_2/D_{2h}$ | 1.58 | 2.37 | A,B | 1.822 | -1905.73 |
| $2^{D_{3h}}_{26} \circ 2^{D_{3h}}_{26} \circ 2^{D_{3h}}_{26}$ | $Si_{330}H_{162}$ | $C_1/C_{2v}$ | 1.58 | 2.96 | A,A | 1.719 | -1907.13 |
| $2^{D_{3h}}_{26} \circ 2^{D_{3h}}_{26} \circ 2^{D_{3h}}_{26} \circ 2^{D_{3h}}_{26}$ | $Si_{430}H_{204}$ | $C_1/C_{2h}$ | 1.58 | 3.92 | A,A | 1.669 | -1907.88 |
| $\alpha\text{-}c-\left(2^{D_{3h}}_{26} \circ 2^{D_{3h}}_{26}\right)_3$ | $Si_{600}H_{252}$ | $C_2/D_{6h}$ | 1.58 | 3.23 | A,B | 1.615 | -1910.36 |
| $\beta\text{-}c-\left(2^{D_{3h}}_{26} \circ 2^{D_{3h}}_{26}\right)_3$ | $Si_{564}H_{252}$ | $C_2/D_{6h}$ | 1.58 | 3.14 | A,B | 1.662 | -1909.13 |
| $\left(2^{T_d}_{28}\right)_n$ family | | | | | | | |
| $2^{T_d}_{28} \circ 2^{T_d}_{28}$ | $Si_{250}H_{132}$ | $C_i/D_{3d}$ | 1.50 | 2.60 | $A_g,A_u$ | 1.980 | -1905.48 |
| $2^{T_d}_{28} \circ 2^{T_d}_{28} \circ 2^{T_d}_{28}$ | $Si_{360}H_{180}$ | $C_1/C_{2v}$ | 1.50 | 3.23 | A,A | 1.927 | -1906.81 |
| $2^{T_d}_{28} \circ 2^{T_d}_{28} \circ 2^{T_d}_{28} \circ 2^{T_d}_{28}$ | $Si_{470}H_{228}$ | $C_1/C_{2h}$ | 1.50 | 4.16 | A,A | 1.893 | -1907.51 |
| mixed junction family | | | | | | | |
| $2^{I_h}_{20} \bullet 2^{D_{6d}}_{24}$ | $Si_{195}H_{102}$ | $C_1/C_s$ | 1.33 | 2.32 | A,A | 1.895 | -1906.10 |
| $2^{I_h}_{20} \bullet 2^{D_{3h}}_{26}$ | $Si_{205}H_{108}$ | $C_s$ | 1.33 | 2.38 | A",A' | 1.930 | -1905.88 |
| $2^{I_h}_{20} \bullet 2^{T_d}_{28}$ | $Si_{215}H_{114}$ | $C_1/C_s$ | 1.33 | 2.41 | A,A | 1.959 | -1905.63 |
| $2^{D_{6d}}_{24} \circ 2^{D_{3h}}_{26}$ | $Si_{220}H_{114}$ | $C_1/C_{2v}$ | 1.48 | 2.25 | A,A | 1.825 | -1906.11 |
| $2^{D_{6d}}_{24} \circ 2^{T_d}_{28}$ | $Si_{230}H_{120}$ | $C_1/C_{3v}$ | 1.48 | 2.36 | A,A | 1.913 | -1905.93 |
| $2^{D_{3h}}_{26} \circ 2^{T_d}_{28}$ | $Si_{240}H_{126}$ | $C_1/C_s$ | 1.58 | 2.48 | A,A | 1.894 | -1905.64 |

[a] The structures are often slightly distorted from a higher point group, which is listed after a slash.

[b] The AM1 values with the correction of 4.91 eV subtracted.

Comment on the three isomers of $3^{I_h}_{20} \bullet 3^{I_h}_{20}$

The formation of the linear $2^{I_h}_{20} \bullet 3^{I_h}_{20}$ and $2^{I_h}_{20} \bullet 3^{I_h}_{20} \bullet 2^{I_h}_{20}$ junctions leads to a significant red energy shift (~0.2 eV) keeping the same incline (Table 2). Depending on the way to produce the $3^{I_h}_{20} \bullet 3^{I_h}_{20}$ junctions through the outer or inner silicon shells of the $3^{I_h}_{20}$, the resulting band gaps differ up to 0.112 eV. The shortest γ-$3^{I_h}_{20} \bullet 3^{I_h}_{20}$ junction with the interface through the inner silicon layers reveals the largest band gap. The other two types (α and β), having the outer silicon layers with the same length but containing a different number of silicon atoms at the interface, show a decrease of the band gap up to 6.562 and 6.453 eV respectively. The band gap of the system with a cavity between the two $3^{I_h}_{20}$ parts (α-$3^{I_h}_{20} \bullet 3^{I_h}_{20}$) is similar to that of the shortest γ-$3^{I_h}_{20} \bullet 3^{I_h}_{20}$ system. The similarity of the band gaps for the systems of different length destroys the typical QCE of the linear $3^{I_h}_{20} \bullet 3^{I_h}_{20}$ systems, making it polysemantic.

References

1. Scholes, G. D. & Rumbles, G. Excitons in nanoscale systems. *Nature Mater.* **5**, 683-696 (2006).
2. Wang, Y., Schmidt, V., Senz, S. & Gösele, U. Epitaxial growth of silicon nanowires using an aluminium catalyst. *Nature Nanotech.* **1**, 186-189 (2006).
3. McAlpine, M. C., Ahmad, H., Wang, D. & Heath, J. R. Highly ordered nanowire arrays on plastic substrates for ultrasensitive flexible chemical sensors, *Nature Mater.* 6, 379-384 (2007).
4. Goswami, S. *et al.* Controllable valley splitting in silicon quantum devices. *Nature Phys.* **3**, 41-45 (2007).
5. Avramov P.V., Chernozatonskii L.A., B. Sorokin P.B., Gordon M.S. Multiterminal Nanowire Junctions of Silicon: A Theoretical Prediction of Atomic Structure and Electronic Properties, *Nano. Lett.* 7, 2063-2067 (2007).
6. Nishiguchi, K., Zhao, X. & Oda, S. Nanocrystalline silicon electron emitter with a high efficiency enhanced by a planarization technique. *J. Appl. Phys.* **92**, 2748-2757 (2002).
7. Cullis, A. G. & Canham, L. T. Visible light emission due to quantum size effects in highly porous crystalline silicon. *Nature* **353**, 335-337 (1991).
8. Cullis, A. G., Canham, L. T. & Calcott, P. D. J. The structural and luminescence properties of porous silicon. *J. Appl. Phys.* **82**, 909-965 (1997).
9. Wolkin, M. V., Jorne, J., Fauchet, P. M., Allan, G., & Delerue, C. Electronic states and luminescence in porous silicon quantum dots: the role of oxygen. *Phys. Rev. Lett.* **82**, 197-200 (1999).
10. Wilcoxon, J. P., Samara, G. A. & Provencio, P. N. Optical and electronic properties of Si nanoclusters synthesized in inverse micelles. *Phys. Rev. B* **60**, 2704-2714 (1999).
11. Zhao, Y. & Yakobson, B. What is the Ground-State Structure of the Thinnest Si Nanowires? *Phys. Rev. Lett.* **91**, 035501 (2003).
12. Filonov, A. B., Ossicini, S., Bassani, F. & d'Avitaya, F. A. Effect of oxygen on the optical properties of small silicon pyramidal clusters. *Phys. Rev. B* **65**, 195317 (2002).
13. Zhao, Y.; Kim, Y.-H.; Du, M.-H. & Zhang, S. B. First-Principles Prediction of Icosahedral Quantum Dots for Tetravalent Semiconductors. *Phys. Rev. Lett.* **93**, 015502 (2004).
14. Caldas, M. J. Si Nanoparticles as a Model for Porous Si. *Phys. Stat. Sol. (B)*, **217**, 641-663 (2000).
15. Garoufalis, C. S., Zdetsis, A. D. & Grimme, S. High Level Ab Initio Calculations of the Optical Gap of Small Silicon Quantum Dots. *Phys. Rev. Lett.* **87**, 276402 (2001).
16. Avramov, P. V. *et al.* Density-functional theory study of the electronic structure of thin $Si/SiO_2$ quantum nanodots and nanowires. *Phys. Rev. B* **75**, 205427 (2007).
17. Goldberg, M. A class of multi-symmetric polyhedra. *Tohoku Math. J.* **43**, 104-108 (1937).
18. Nishio, K., Morishita, T., Shinoda, W. & Mikami, M. Molecular dynamics simulations of self-organized polyicosahedral Si nanowire, *J. Chem. Phys.* **125**, 074712 (2006).
19. Dewar, M. J. S., Zoebisch, E. G., Healy, E. F. & Stewart, J. J. P. Development and use of quantum mechanical molecular models. 76. AM1: a new general purpose quantum mechanical molecular model. *J. Am. Chem. Soc.* **107**, 3902-3909 (1985).
20. Stewart, J. J. P. Optimization of parameters for semiempirical methods I. Method. *J. Comput. Chem.* **10**, 209-220 (1989).
21. Mazzone, A. M. Properties of pure and compound clusters of Si, Ge, and Pb, *Phys. Rev. B* **54**, 5970-5977 (1996).
23. Sze S. M. *The Physics of Semiconductor Devices,* Wiley, New York (1969).